\documentclass[pre,twocolumn,amsmath,floatfix]{revtex4}
\usepackage{xcolor}
\usepackage{graphicx}
\usepackage[colorlinks=true,linkcolor=blue,citecolor=blue]{hyperref}

\begin{document}
\title{Channel Noise Effects on Neural Synchronization}
\author{Brenton Maisel}
\email{bmaisel@ucsd.edu}
\affiliation{Department of Chemistry and Biochemistry, and BioCircuits Institute, University of California San Diego, La Jolla, California 92093-0340, USA}
\author{Katja Lindenberg}
\affiliation{Department of Chemistry and Biochemistry, and BioCircuits Institute, University of California San Diego, La Jolla, California 92093-0340, USA}

\begin{abstract}
Synchronization in neural networks is believed to be linked to cognitive processes, while abnormal synchronization has been associated with disorders such as epilepsy and schizophrenia. We examine the synchronization of small Hodgkin-Huxley neuronal networks. The principal features of Hodgkin-Huxley neurons are protein channels in the neural membrane that transition between open and closed states with voltage dependent rate constants. The standard assumption of infinitely many channels neglects the fact that real neurons have finitely many channels, which leads to fluctuations in the membrane voltage and modifies neuronal spike times. These fluctuations are referred to as channel noise. We demonstrate that regardless of channel noise magnitude, neurons in the network reach a steady state synchronization level dependent only on the number of neurons in the network, equivalent to the steady state level of uncoupled Poisson neurons. The channel noise only affects the time to reach the steady state synchronization level.


\end{abstract}

\maketitle

\section{Introduction}
\label{intro}
The human brain is estimated to contain over $100$ billion neurons, with each neuron connected to approximately $10^4$ other  neurons~\cite{gerstner2014neuronal}. Neurons influence each other through excitatory and inhibitory synaptic connections and, as a result, neurons in a network are rhythmically activated and inhibited through their synaptic connectivity~\cite{buzsaki2006rhythms}. Synchronized interactions across brain regions have been proposed to underly a wide variety of cognitive functions~\cite{van2012neural,fries2005mechanism,fell2011role,schnitzler2005normal,singer1993synchronization}. For example, when monkeys learn categorical information (e.g., how to distinguish between groups of negative and positive objects), experiments have shown increased neural synchronization between the prefrontal cortex and the striatum~\cite{antzoulatos2014increases}.

In addition to the cognitive roles of synchronization, abnormal synchronization has been linked to a number of brain disorders such as epilepsy, schizophrenia, Alzheimer's disease, and Parkinson's disease~\cite{abuhassan2014compensating,hammond2007pathological,uhlhaas2006neural}.
For example, epilepsy has commonly been associated with excessive synchronization of neural populations~\cite{jiruska2013synchronization,uhlhaas2006neural}, whereas schizophrenia has been associated with impaired neural synchronization~\cite{spencer2003abnormal,uhlhaas2010abnormal}. In combination with cognitive functions, it is clear that the balance between synchronized and asynchronized neural oscillations plays an important role in healthy brain activity.

In order to model neuronal synchronization, many studies have focused on networks of Hodgkin-Huxley neurons. One defining property of the Hodgkin-Huxley neuron model that allows for action potential generation is the existence of sodium and potassium channels in the neuron membrane that transition between open and closed states with voltage-dependent rate constants. Each channel is composed of four gates: the sodium channel is composed of three activating gates (known as type $m$ gates) and one inactivating gate (known as a type $h$ gate), and the potassium channel is composed of four activating gates of type $n$. At rest, the activating gates are closed and the inactivating gate is open, but as a neuron receives synaptic input from other neurons, the membrane voltage rises, causing the activating gates to open. This starts the depolarization of the membrane potential. When the voltage is sufficiently high, the sodium inactivating gate closes while the potassium gates remain open, which repolarizes the membrane potential. The dynamics of the $n$ gates and $m$ gates are similar, but the $n$ gate dynamics evolve on a slower time scale~\cite{wells2007introduction}.

However, while the Hodgkin-Huxley model assumes that there are infinitely many channels so that fluctuations in the number of open channels remain undetected, real neurons have only finitely many channels and hence experience intrinsic noise through the stochastic opening and closing of sodium and potassium protein channels in the neural membrane \cite{chow1996spontaneous,white2000channel,hille2001ionic}. This stochasticity leads to fluctuations in the membrane potential which can alter spike timing of neurons. We refer to this stochasticity as channel noise, and its role in neural synchronization is the focus of our paper.

In this study, we investigate the synchronization of a small network of noisy Hodgkin-Huxley neurons. We select this neuron for study because of its close connection to biological reality and its ability to reproduce almost all single-neuron properties~\cite{hodgkin1952quantitative,bukoski2015channel,izhikevich2004model}. A channel can only conduct when it is considered open, and a channel is considered open when all the gates within the channel are open. The most direct approach to modeling the opening and closing of channels is referred to as the Markov Chain model. This model is consistent with the Hodgkin-Huxley model in the statistics of interspike intervals and is computationally much less intensive. In the Markov Chain model, each of $N$ channels of a particular type functions as a Markov process, transitioning between open and closed states with voltage-dependent rate constant, independently of other same-type channels. At each time step, one then determines the fraction of open channels of a particular type. These fractions are then used explicitly in the Hodgkin-Huxley voltage equation to model fluctuations in the membrane voltage due to channel noise. Typically, the Markov Chain model is simulated using a Gillespie algorithm~\cite{gillespie1977exact,chow1996spontaneous,skaugen1979firing}. However, while computationally more modest than the Hodgkin-Huxley model that it is chosen to exemplify, simulating such a Markov Chain is still computationally exhaustive as the necessary time step for simulations is of order $N^{-1}$. As an alternative method to account for perturbations in the Hodgkin-Huxley model, some studies have added an external perturbation to the Hodgkin-Huxley equations to assess the role of noise in synchronization~\cite{borges2016effects,lameu2012suppression,popovych2013self}. This method, however, lacks justification that it accurately models the stochastic opening and closing of channels. 

Fox and Lu derived a set of stochastic differential equations which approximate the behavior of the Markov Chain model, and these equations will be used in this paper \cite{fox1994emergent}.Their model was developed by using a system size expansion applied to the Markov Chain version of the Hodgkin-Huxley equations. As a result, each dynamical variable in their system of equations represents the fraction of ion channels in a specific configuration. The stochastic equations of Fox and Lu do not modify the deterministic structure of the Hodgkin-Huxley equations, and they include stochastic perturbations that account for the opening and closing of channels. We refer to these stochastic perturbations as channel noise to be consistent with previous literature~\cite{goldwyn2011stochastic,goldwyn2011and}. Results of numerical simulations of this model agree remarkably well with the dynamical behavior predicted by the Markov Chain model of the channel states~\cite{goldwyn2011stochastic,goldwyn2011and}. To summarize, we have discussed three models: the original Hodgkin-Huxley model of a neuron, the Markov Chain model which models stochasticity in the channels transitioning randomly between open and closed states, and the Fox and Lu model which we will use due to its highly accurate approximation of the Markov Chain model but with a much faster computational time. Our goal is to understand how channel noise affects the synchronicity of neurons in a network, and we will use the Fox and Lu model to do so.

Our paper is organized as follows: In Sec.~\ref{model} we present a mathematical description of the stochastic Hodgkin-Huxley neuron with a synaptic connectivity term. Then in Sec.~\ref{results} we derive a formula to estimate the degree of synchronization based on the size of the neural network, and we demonstrate that channel noise causes the synchronization of the neural network to behave as independent Poisson neurons. We then conclude with some closing remarks.

\section{Model}
\label{model}
One of the most important models in computational neurosciences is the Hodgkin-Huxley neuronal model of a squid  axon~\cite{hodgkin1949effect,hodgkin1952quantitative}. The deterministic dynamics of the Hodgkin-Huxley model are given by the following set of differential equations:

\begin{eqnarray}
\label{HH}
C\dot{V} &=& I(t) - \bar{g}_{Na}m^3h(V-E_{Na}) \nonumber \\ &-& \bar{g}_Kn^4(V-E_K) - \bar{g}_L (V-E_L) \nonumber\\
\dot{n} &=& \alpha_n(V) (1-n) - \beta_n(V)n \\
\dot{m} &=& \alpha_m(V) (1-m) - \beta_m(V)m \nonumber\\
\dot{h} &=& \alpha_h(V) (1-h) - \beta_h(V)h, \nonumber
\end{eqnarray}

\noindent  where $V(t)$ is the time-dependent voltage  due to the charge difference inside and outside the membrane that surrounds a neuron, and $I(t)$ is the input current to the neuron from all sources. The membrane potential-dependent coefficients are chosen to be given by

\begin{eqnarray}
\alpha_n(V) &=& \frac{0.01V+0.55}{1-\exp[-0.1V-5.5]} \nonumber\\
\beta_n(V) &=& 0.125\exp[-(V+65)/80] \nonumber\\
\alpha_m(V) &=& \frac{0.1V+4}{1-\exp[-0.1V-4]} \nonumber\\
\beta_m(V) &=& 4\exp[-(V+65)/18] \nonumber\\
\alpha_h(V) &=& 0.07\exp[-(V+65)/20] \nonumber\\
\beta_h(V) &=& \frac{1}{1+\exp[-0.1V-3.5] }. \nonumber
  \end{eqnarray}

\noindent The numerical coefficients are those used in the original Hodgkin-Huxley paper.  The values of the parameters (determined experimentally) along with definitions are found in Table~\ref{parameters}~\cite{hodgkin1952quantitative}. The original parameter choices were made so that the resting potential is $0$mV;  most literature using the Hodgkin-Huxley model use these same values.
Our resting potential is $-70$mV, which is a typical voltage across an animal cell membrane, and is easily obtained by shifting parameter values. It is also more convenient for our further calculations. 

\begin{table}[ht]
\caption{Parameter values used for simulation of the Hodgkin-Huxley model.} 
\centering 
\begin{tabular}{c c c} 
\hline\hline 
Parameter & Definition &Value \\[0.5ex] 
\hline 
$C$ & membrane capacitance&$1 \mu F /cm^2$  \\
$E_{Na}$ &sodium reversal potential& $50 mV$ \\ 
$E_K$ &potassium reversal potential& $-77 mV$\\
$E_L$ &leak reversal potential& $-54.4 mV$\\
$\bar{g}_{Na}$ &maximal sodium conductance& $120 mS/cm^2$\\
$\bar{g}_K$ &maximal potassium conductance &$36 mS/cm^2$\\
$\bar{g}_L$ &maximal leak conductance& $0.3 mS/cm^2$\\
\hline 
\end{tabular}
\label{parameters} 
\end{table}

The most direct method of working with the associated Markov Chain model is to take a population of ion channels and a small time step $dt$, and then calculate the probability that each channel flipped from its current state to another state during this time interval. Although this method works, it is extraordinarily slow when the number of ion channels is large~\cite{mino2002comparison}. To counter this problem, we use the Fox and Lu system size expansion, which is a set of stochastic differential equations that replicates the behavior of the Markov Chain model with high accuracy and less computational exhaustion~\cite{fox1994emergent,goldwyn2011stochastic,goldwyn2011and}. The Fox and Lu system size expansion is given by the following set of stochastic differential equations:

\begin{eqnarray}
\label{SHH}
C\dot{V} &=& I_{inj}(t)+I_{syn}(t) - \bar{g}_{Na}y_{31}(V-E_{Na}) \nonumber \\ &-&\bar{g}_Kx_4(V-E_K) - \bar{g}_L (V-E_L)\nonumber\\ \nonumber\\ \nonumber\\
\dot{\mathbf{x}} &=& A_K(V)\mathbf{x}+\frac{1}{\sqrt{N_{K}}}S_K(V,\mathbf{x}) \xi_K \\
\dot{\mathbf{y}} &=& A_{Na}(V)\mathbf{y}+\frac{1}{\sqrt{N_{Na}}}S_{Na}(V,\mathbf{y})\xi_{Na} \nonumber ,
\end{eqnarray}

\noindent where the input current $I(t)$ has been decomposed into two contributions. $I_{inj}$ determines whether or not action potentials occur~\cite{luccioli2006dynamical}. Action potentials are variations in the voltage of the neuron membrane; when an action potential is triggered, the membrane potential abruptly shoots upward (fires) and then equally abruptly shoots back downward . The $I_{syn}$ term represents current input from the chemical synapses of other neurons in the network. 
Tthe matrices $A_K$, $A_{Na}$, $S_K$, and $S_{Na}$  are written out in the Appendix. The vector $\mathbf{x}$ is composed of components $x_i$ $(i = 0, 1,2,3,4)$ representing the proportion of potassium channels with $i$ open gates of type $n$. The entries of $\mathbf{y}$ are denoted as $y_{ij}$ $(i=0,1,2,3$ and $j=0,1)$, representing the proportion of sodium channels with $i$ open $m$ subunits and $j$ open subunits of type $h$. While Eq.~(\ref{SHH}) is valid for a large number of channels, it has been shown to be a very accurate representation of the Markov Chain model even for a small number of channels~\cite{goldwyn2011stochastic}.

We are interested in a neural network in which the connections between neurons are unidirectional and the local dynamics are described by the Fox and Lu system size model Eq.~(\ref{SHH}). The $I_{syn}$ term is given by 
\begin{equation*}
I_{syn} = \frac{(V_r-V_i)}{\Omega}\sum_{j=1}^N \epsilon_{ij}s_j,
\end{equation*}
with time-dependent entries defined by the following set of ordinary differential equations: \cite{destexhe1994efficient,borges2016effects,golomb1993dynamics}:
\begin{equation*}
\dot{s}_i = \frac{5(1-s_i)}{1+\exp(-\frac{V_i+3}{8})}-s_i
\end{equation*}

\noindent Here, $V_r$ is the synaptic reversal potential set to $20 mV$~\cite{popovych2013self}, $s_i$ is the post-synaptic potential of neuron $i$, $\epsilon_{ij}$ represents the synaptic coupling strength between the $j^{th}$ presynaptic neuron and the $i^{th}$ postsynaptic neuron, and $\Omega$ is the average number of connections at each synapse, which we take to be $1$. The remaining parameter, $I_{inj}$, determines whether or not action potentials occur~\cite{luccioli2006dynamical}. For small values of $I_{inj}, I_{inj} < 6.27 \mu A/cm^2$, the deterministic Hodgkin-Huxley model resides in a silent regime in that action potentials are not generated. When the injected current is greater than $9.78 \mu A/cm^2$, the deterministic Hodgkin-Huxley model enters the repetitive firing regime, that is, action potentials are generated. Between these values, known as the excitable region, the model shows bistability between silence and repetitive firing.

In this paper we are interested in how channel noise affects the synchronicity of neuron spiking at different firing rates. In order to quantitatively study neural synchronicity, we use the order parameter $R$ defined as

\begin{equation}
\label{OP}
R(t) = \left| \frac{1}{N}\sum_{j=1}^N \exp(i \theta_j) \right| ,
\end{equation}

\noindent where $\theta_j(t)$ is the phase of the $jth$ neuron defined by~\cite{acebron2005kuramoto,borges2016effects}
\begin{equation}
\theta_j(t)= 2\pi m + 2\pi \frac{t-t_{j,m}}{t_{j,m+1}-t_{j,m}},
\label{RRR}
\end{equation}

\noindent and $\theta_j(t)=0$ for $t < t_{j,1}$. In this equation, $t_{j,m}$ denotes the time when neuron $j$ emits spike $m$ $(m=0,1\cdots)$. Equation~(\ref{OP}) is designed in such a way that the first spike begins at $\theta = 0$ and the phase increases linearly until the next spike occurs at $\theta = 2\pi$. If all neurons are completely synchronized, then $\theta_1(t) = \dots = \theta_N(t)$, and hence 

\begin{equation*}
R = \left| \frac{1}{N}\sum_{j=1}^N \exp(i \theta_j) \right|= \left| \frac{1}{N}N\exp(i \theta_1) \right| = \left| \exp(i \theta_1) \right| = 1. 
\end{equation*}

\noindent Therefore $R$ values are closer to unity when neurons have more synchronized spike times.

\section{Synchronization}
\label{results}
The stochastic differential equations Eq.~(\ref{SHH}) can not be handled analytically, and so we must continue our analysis on the basis of numerical simulations. We used the Euler-Maruyama method~\cite{higham2001algorithmic,gard1988introduction} with time step $\Delta t = 10 \mu s$. Unless noted otherwise, initially each neuron in the network was assumed to be in the resting state.

\subsection{Frequency-Current Relationship}
\label{FCRel}
\begin{figure*}
    \centering
    \includegraphics[width=0.55\textwidth]{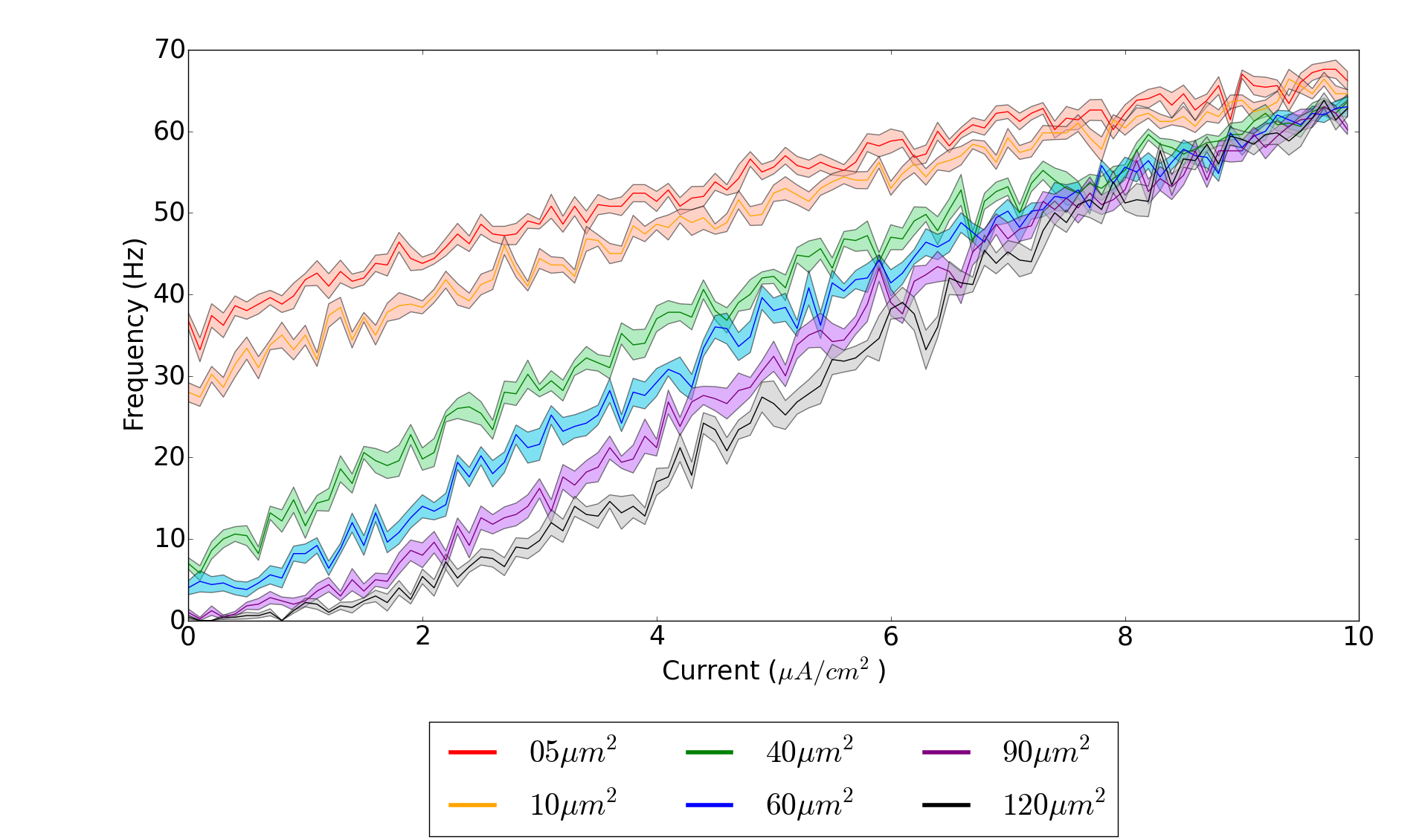} 
    \caption{Relationship between input current and firing frequency for different membrane areas. Solid lines show the mean     firing frequency averaged over $50$ simulations. Shaded areas show one standard deviation of firing frequency from the mean.}
    \label{fI}
\end{figure*}

To understand how the number of channels affects the firing rate of a Hodgkin-Huxley neuron, we analyze the relationship between firing frequency of a stochastic neuron and the input current. In the squid axon modeled by Hodgkin and Huxley, the ratio of sodium channel density to potassium channel density is approximately $60 \mu m^{-2} / 18 \mu m^{-2}$, and we use these values for our simulations \cite{ferreira1985biophysical}. Defining $A$ to be the membrane area, the total number $N_{Na}$ of sodium channels is proportional to $60\times A$ and the total number $N_K$ of potassium channels is proportional to $18 \times A$. From Eq.~(\ref{SHH}), the parameter $A$ therefore controls the magnitude of fluctuations from the channel noise as $A^{-1/2}$. Therefore, a smaller membrane area results in larger fluctuation magnitude and correspondingly larger membrane area results in smaller fluctuation magnitude. The resulting firing frequencies as a function of input current for different membrane areas are given in Fig.~\ref{fI}.

The results show that in the absence of input current, the size of the membrane area is what primarily determines the rate of spontaneous firing. As the input current increases, the firing rates for all different areas converge towards each other and the firing rate becomes independent of the membrane area. With an increase in current, the neurons enter the repetitive firing regime so fluctuations in the membrane voltages should produce only minor changes in the frequency of spikes. Therefore, channel noise has a larger effect on the firing rate at low input currents, in agreement with previous literature \cite{skaugen1979firing,clay1983relationship,strassberg1993limitations}. 

\subsection{Channel Number Effect on Synchronization}
\begin{figure}
    \centering
    \includegraphics[width=7cm]{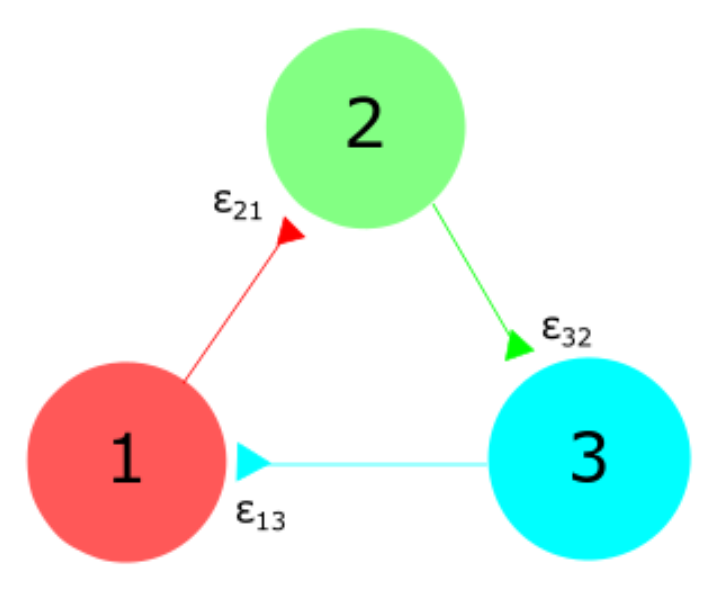}
    \caption{Schematic drawing showing three unidirectionally connected excitatory neurons with coupling strengths     $\epsilon_{13}, \epsilon_{21},$ and $\epsilon_{32}$}.
    \label{netw}
\end{figure}

To understand how the number of channels affects the synchronicity of neurons, we consider a simple three member neural network with unidirectional excitatory connections and local dynamics given by Eq.~\ref{SHH}. Such a system is shown in Fig.~\ref{netw}. In this example we assume the connections to have identical coupling constants $\epsilon_{13} = \epsilon_{21} = \epsilon_{32} = 0.10$ and zero otherwise.

\begin{figure*}
    \centering
    \includegraphics[width=0.75\textwidth]{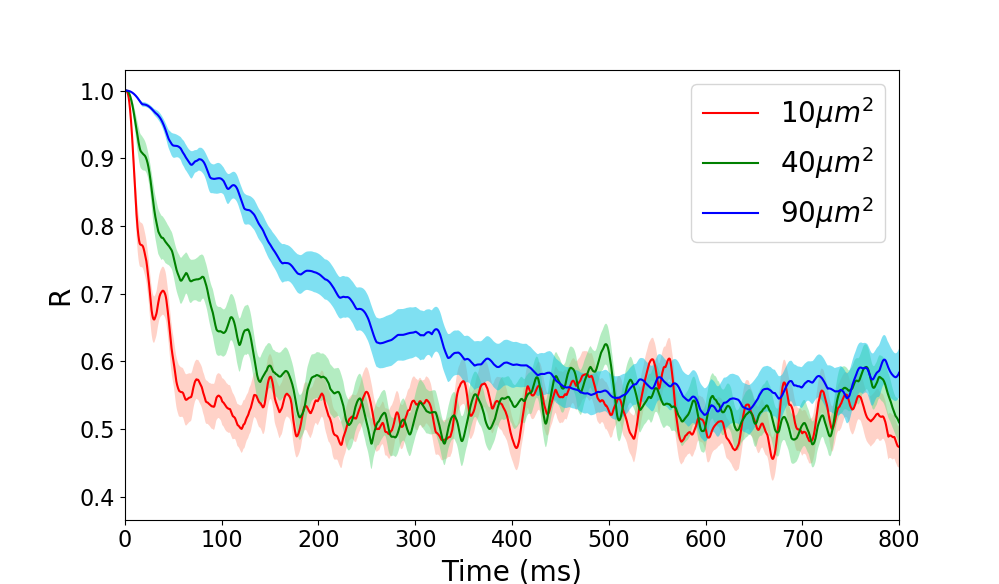}
     \caption{Simulations of Eq.~(\ref{SHH}) using the system shown in Fig.~\ref{netw} for varying membrane areas. Bold lines represent the mean after $50$ simulations while shaded areas show one standard deviation from the mean. Injected current was $10.0 \mu A/cm^2$.}
    \label{DA}
\end{figure*}

Because the neurons all start with the same initial condition, we have that $R(0)=1$ regardless of the value of the membrane area, i.e., the neurons begin completely synchronized. However, as time passes, the degree of synchronization changes. Figure~\ref{DA} shows simulations of the three neuron neural network with different membrane areas. Although the degree of synchronization for smaller area changes more rapidly than in the case of larger membrane area, the degree of synchronization appears to reach the same steady state value and hover around this value. From this simulation, the membrane area only affects the time to reach the steady state synchronization value but not the steady state synchronization value itself. Since changing membrane area changes the firing rate (see Fig.~\ref{fI}), this suggests that there is an inverse relationship between the firing rate and the time it takes to reach a steady state synchronization level (and consequently, a direct relationship between membrane area and time to reach steady state).

\subsection{Comparison to Independent Poisson Neurons}
\label{derivation}
We wish to compare the observed steady state value in Fig.~\ref{DA} with the predicted value of Eq.~(\ref{OP}) when the neurons are completely independent Poisson neurons, that is, neurons whose firing frequency is a Poisson process independent of other neurons. We will then later extend this result to a larger network. In large network modeling, the distribution of spike times is highly irregular, and modeling neurons in the network as a Poisson process is widely used \cite{zador1996integrate,gerstner2014neuronal,izhikevich2003simple}. As shown in Fig.~\ref{fI}, for a given input current, channel noise affects the firing rate of a neuron. We can therefore understand the relationship between channel noise and synchronization by studying the relationship between synchronization and firing rate.

To compare the steady state synchronization level observed in simulations with that of independent Poisson neurons, we proceed to calculate the expectation of our order parameter $R(t)$, which we denote as $\langle R(t) \rangle$. For simplicity, let us initially assume we have two neurons $N_1$ and $N_2$ whose spike times follow a Poisson distribution with rate parameter $\lambda$, and

\begin{eqnarray}
\langle R(t)\rangle = \langle\sqrt{R(t)^2}\rangle &=& \frac{1}{2}\langle\sqrt{|e^{i\theta_1(t)} + e^{i\theta_2(t)}|^2} \rangle \nonumber \\
&=& \frac{1}{2}\langle\sqrt{2+2\cos\left(\theta_1(t)-\theta_2(t)\right)} \rangle. \nonumber
\end{eqnarray}

\noindent To compute this expectation, we are required to find the joint distribution of $\theta_1$ and $\theta_2$. Since we are assuming that the Poisson neurons are independent, we only need to find the density function of $\theta_1$. 

From the definition of $\theta$, only the random term $(t-t_{m})/(t_{m+1}-t_{m})$ has any bearing on the order parameter (cf. Eq.~(\ref{RRR})). Therefore, to understand the distribution of $\theta$, we only need to understand the distribution of $(t-t_{m})/(t_{m+1}-t_{m})$. Since we observe a steady state of synchronization in the simulations after time passes, consider a time $t$ where $t$ is large enough so that at least one spike has occurred before time $t$. Let $X$ be a random variable describing the length of time between our time $t$ and the time of the most recent spike before time $t$. Moreover, let $Y$ be the length of time between time $t$ and the next spike after time $t$. Notice that the ratio $X/(X+Y)$ corresponds directly to the $(t-t_{m})/(t_{m+1}-t_{m})$ term in the definition of $\theta$ in Eq.~(\ref{RRR}). We need to add one constraint to $X$. Because $X$ is the length of time between $t$ and the previous spike, the maximum value $X$ can take is $t$ (otherwise the previous spike had to occur before time $0$, which is not relevant to our calculations). Therefore, $(t-t_{m})/(t_{m+1}-t_{m})$ can be modeled by $\min \left\lbrace X,t\right\rbrace/(\min \left\lbrace X,t\right\rbrace + Y)$. Since the spike times of the neurons are being treated as Poisson processes, the wait time between spikes is an exponential distribution with rate parameter $\lambda$. Therefore, we have that $Y$ has an exponential distribution with rate parameter $\lambda$, $X$ also has an exponential distribution with rate parameter $\lambda$, and $X$ and $Y$ are independent of each other. 

Exponential distributions have the scaling property, which means that if $X$ has an exponential distribution with rate $\lambda$, then $X$ has the same distribution as $\lambda^{-1}\xi$ where $\xi$ is an exponential random variable with rate $1$. Thus, $Y$ also follows the distribution $\lambda^{-1}\eta$ where $\eta$ is an exponential random variable with rate $1$. Putting everything together, we have:

\begin{eqnarray}
\frac{t-t_{m}}{t_{m+1}-t_{m}} \sim \frac{\min \left\lbrace X,t\right\rbrace}{\min \left\lbrace X,t\right\rbrace + Y} &=& \frac{\min \left\lbrace \lambda^{-1}\xi,t\right\rbrace}{\min \left\lbrace \lambda^{-1}\xi,t\right\rbrace + \lambda^{-1}\eta} \nonumber \\
&=& \frac{\min \left\lbrace \xi,\lambda t\right\rbrace}{\min \left\lbrace \xi,\lambda t\right\rbrace + \eta}. \nonumber
\end{eqnarray}

\noindent We next assume that $\lambda t$ is large enough so that min$\{\xi,\lambda t\} = \xi$. We can then approximate the distribution of $\theta$ to be that of $\frac{\xi}{\xi + \eta}$ where $\xi$ and $\eta$ are independent exponential random variables with rate parameter $1$. To find the density function of $\theta$, we only need to determine the density function for $\frac{\xi}{\xi + \eta}$. This is known to be a uniform distribution, but we will show this for the purpose of completeness~\cite{sheldon2002first}. To calculate this density function, we will find the cumulative distribution function and take its derivative. Note that $0 < \frac{\xi}{\xi + \eta} < 1$, so we can pick an arbitrary value $a \in (0,1)$ to use for calculating the distribution function. We use $f_{\xi}(x)$ to mean the density function of $\xi$ in the derivation. We have:

\begin{eqnarray}
P\left(\frac{\xi}{\xi+\eta} \leq a \right) &=& P\left(\frac{\xi + \eta}{\xi} \geq \frac{1}{a} \right) \nonumber \\
&=& P\left(\eta \geq \xi\left(\frac{1}{a}-1\right) \right) \nonumber \\
&=& \int_0^\infty P\left(\eta \geq s\left(\frac{1}{a}-1\right) \right) f_\xi(s) ds  \nonumber \\
&=& \int_0^\infty e^{-s\left(\frac{1}{a}-1\right)}e^{-s} ds \nonumber \\
&=& \int_0^\infty e^{-\frac{s}{a}} ds \nonumber \\ \nonumber \\ 
&=& a. \nonumber
\end{eqnarray}

This means that $\frac{\xi}{\xi + \eta}$ has the same distribution as a uniform random variable on $(0,1)$, as expected. Consequently, the density function of $\theta$ is just $1$. Returning to our calculation of the expectation of synchronization:

\begin{eqnarray}
\langle R(t) \rangle &=& \frac{1}{2}\langle\sqrt{2+2\cos\left(\theta_1(t)-\theta_2(t)\right)}\,\, \rangle  \nonumber \\
&=& \frac{1}{2}\int_0^1\int_0^1 \sqrt{2+2\cos\left(2\pi x_1-2\pi x_2 \right)} dx_1dx_2. \nonumber\\
\end{eqnarray}

This double integral can be solved exactly by utilizing a simple substitution and recognizing that we are integrating over one period of the cosine function, 

\begin{eqnarray}
\langle R(t) \rangle &=& \frac{1}{2}\int_0^1\int_0^1 \sqrt{2+2\cos\left(2\pi x_1-2\pi x_2 \right)} dx_1dx_2 \nonumber \\ 
&=& \frac{1}{8\pi^2}\int_0^{2\pi}\int_0^{2\pi} \sqrt{2+2\cos\left(x_1 - x_2 \right)} dx_1dx_2 \nonumber\\
&=& \frac{1}{4\pi}\int_0^{2\pi}\sqrt{2+2\cos\left(x_1 \right)} dx_1 \nonumber \\
&=& \frac{2}{\pi}. \nonumber
\end{eqnarray}

Remarkably, we have shown that in the long-time limit of weakly coupled neurons, the expected steady state synchronization level has no dependence on the firing rate (membrane area) of the neurons. We will confirm, as already implicit, that the steady state synchronization depends only on the number of neurons in the network. In order to obtain a solution for the two-neural-network system, we needed to make a few assumptions: (a) the neurons in the network were weakly coupled to approximate them as independent of each other, and (b) the quantity $\lambda t$ is sufficiently large. Expanding on the second point, recall that we approximated $\min \left\lbrace \xi,\lambda t\right\rbrace$ by $\xi$. Since $\xi$ is exponentially distributed with rate parameter $1$, then $P\left(\xi \leq \lambda t \right) = 1-\exp(-\lambda t)$. Because of the exponential decay dependence on $\lambda t$, this means that $\lambda t$ does not have to be very large before one can approximate $\min \left\lbrace \xi,\lambda t\right\rbrace$ by $\xi$ with high probability. With the application to neurons, this implies that when the firing rate of neurons is higher (small membrane area), we should expect less time to reach a steady state synchronization level. Conversely, with a lower firing rate (large membrane area), we should expect a longer time to reach steady state synchronization. The high and low firing rates correspond with lower and higher channel noise magnitude respectively. Therefore, this derivation provides justification for the observation in Fig.~\ref{DA} that larger channel noise magnitude in neurons results in faster desynchronization. 

To expand the result above to a larger neuron network, note that the only place where changes will occur is in the term under the radical. That term results from simplifying $\sqrt{R(t)^2}$, and a formula for $\langle R(t)\rangle$, as follows:

\begin{equation}
\label{GS}
\langle R(t) \rangle = \frac{1}{N}\int_0^1 \sqrt{N + \sum_{\substack{j,k=1 \\ j\neq k}}^N\cos\left(2\pi x_j-2\pi x_k \right)}dx_1\dots dx_N.
\end{equation}

\noindent Unlike the two neural network case, higher dimensional cases of Eq.~\ref{GS} must be evaluated numerically. Numerical estimations for the steady state synchronization predicted by Eq.~( \ref{GS}) for different numbers of neurons $N$ can be found in Table~\ref{tabl}.

\begin{table}[ht]
\caption{Steady state synchronization values estimated from Monte Carlo simulations of Eq.~(\ref{GS})} 
\centering 
\begin{tabular}{c c} 
\hline\hline 
Number of Neurons (N) & Steady State Synchronization Value \\[0.5ex] 
\hline 
2 & 0.636\\
3 & 0.525 \\ 
4 & 0.450 \\
\hline 
\end{tabular}
\label{tabl} 
\end{table}

\begin{figure*}
    \centering
    \includegraphics[width=0.75\textwidth]{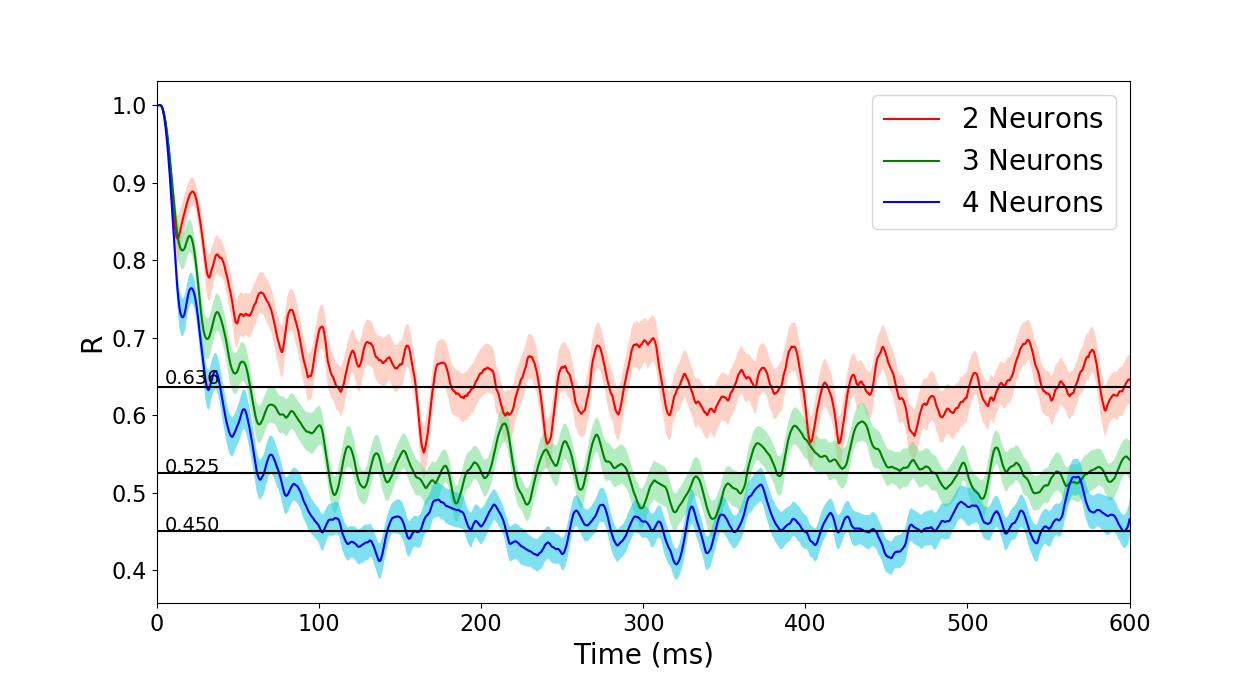}
   \caption{(Color Online) Simulations showing synchronization parameter behavior for neuron networks of $2, 3,$ and $4$ neurons. Straight lines show estimated steady state
    synchronization values as determined by Eq.~\ref{GS}. Each neuron in the network had an area of $40 \mu m^2$ and an injected current of $8.0 \mu A/cm^2$. Bold lines represent the mean after $100$ simulations while shaded areas show one standard deviation from the mean.}
    \label{SteadyState}
\end{figure*}

The comparison between the values obtained above and numerical simulations is shown in Fig.~\ref{SteadyState}. As shown in Fig.~\ref{SteadyState}, our estimations of the steady state synchronization values as determined by Eq.~(\ref{GS} ) is quite accurate, demonstrating that in the presence of channel noise, the synchronization of our neural network behaves just as that of independent Poisson neurons.

\begin{figure}
    \centering
    \includegraphics[width=9cm]{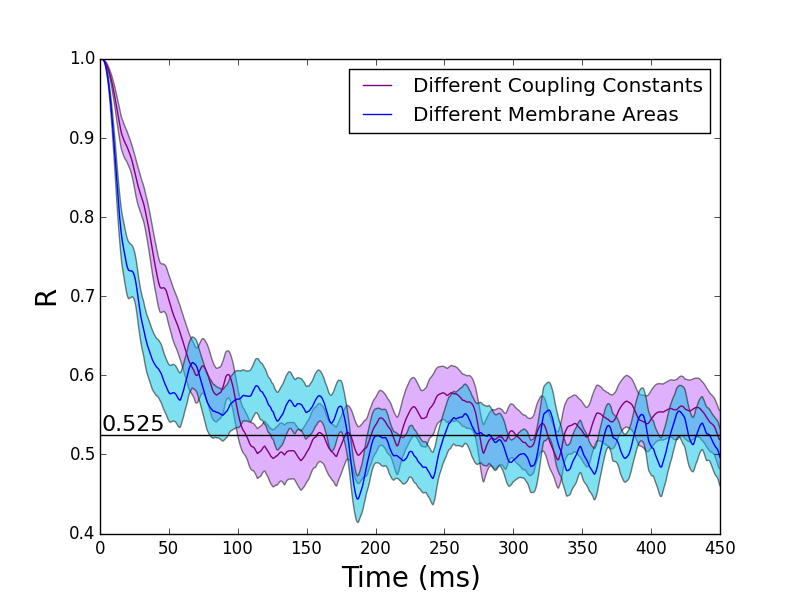}
    \caption{(Color Online) Simulations of the network shown in Fig.~\ref{netw}. For the coupling constant plot, we use the parameters $    \epsilon_{21} = 10, \epsilon_{32} = 10^{-1},$ and $\epsilon_{13} = 10^{-3}$ with each neuron having a membrane area of $40 \mu m^2$. For the membrane area plot, we considered each neuron having a different membrane area of $10 \mu m^2$, $40 \mu m^2$, and $90 \mu m^2$. All coupling constants for the different membrane area simulation were set to $0.10$.}
    \label{AP}
\end{figure}

It is also worth noting that based on our results, the connectivity of neurons in the network has no bearing on the steady state synchronization as long as the coupling is weak. To emphasize, the importance of synchronization of the network is not the connectivity of neurons, or the initial states of the neurons, but only the number of neurons involved in the network. The role of noise is to help change the rate at which the network reaches a steady state synchronization level but does not appear to change the level itself. To generalize this result a bit further, we consider two additional cases. We will examine what happens if we change the connectivity strengths so that the connections are no longer equal, and we will change the areas so each one is affected by a different magnitude of channel noise. These results are shown in Fig~\ref{AP}.

For the case of unequal coupling constants, we considered the network in Fig.~\ref{netw} with all neurons having a membrane area of $40 \mu m^2$ and coupling constants of $\epsilon_{21} = 10, \epsilon_{32} = 10^{-1},$ and $\epsilon_{13} = 10^{-3}$. For the case of various channel noise magnitudes, we considered all synaptic connections to be $0.10$. Neurons 1, 2, and 3 had membrane area of $10 \mu m^2$, $40 \mu m^2$, and $90 \mu m^2$ respectively. In both cases, the expected steady state deviation again approaches that of independent Poisson neurons, and this was observed over a wide range of values. This result suggests that the resulting formula for the expected steady state synchronization obtained for independent Poisson neurons is applicable not just to Poisson neurons or a Hodgkin-Huxley network with equal coupling constants, but generalizes to Hodgkin-Huxley networks that have unequal coupling and differing membrane areas.

\subsection{Large Membrane Area}
\label{LMA}
\begin{figure*}
    \centering
    \includegraphics[width=0.75\textwidth]{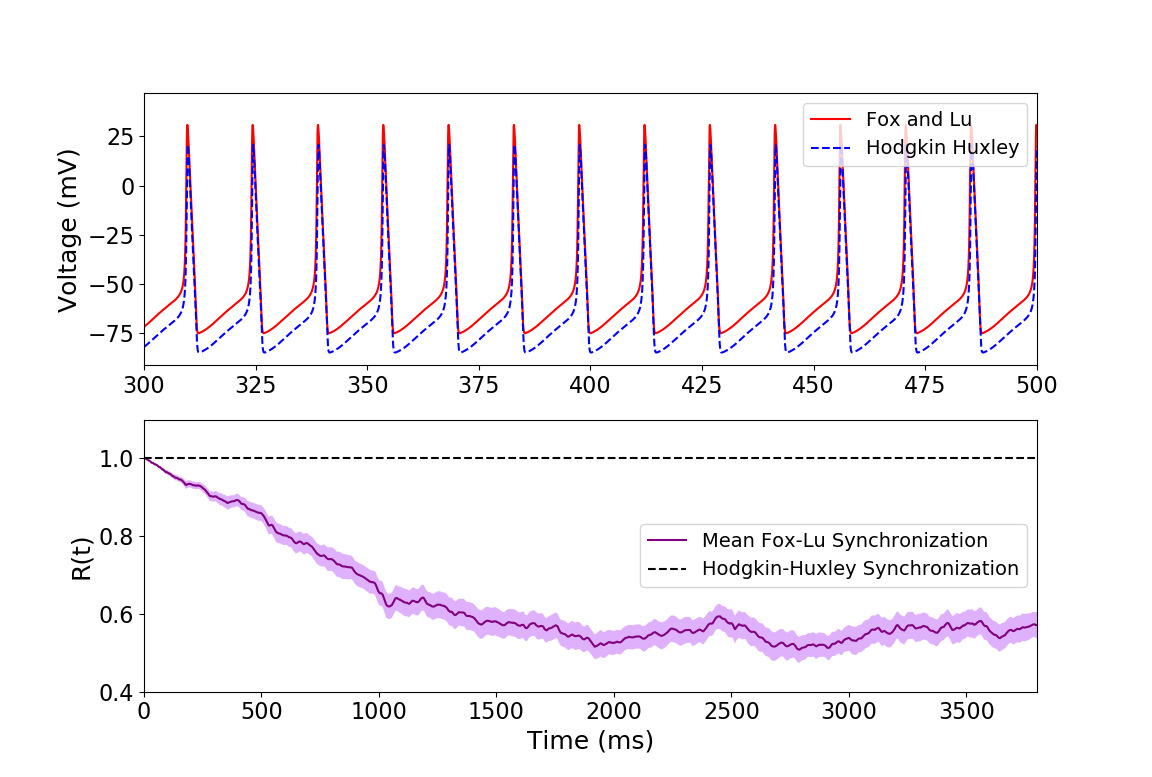}
     \caption{(Color Online) (a) Comparison of membrane voltages for the deterministic Hodgkin-Huxley model and Fox and Lu model with membrane area of $300 \mu m^2$. Both      plots were conducted with input current of $10.0 \mu A/cm^2$. Membrane voltage for the Hodgkin-Huxley model was offset by $10$ mV for clarity. (b) Order parameter for the system shown in Fig.~\ref{netw} where all three neurons are either deterministic or stochastic with an area of $300 \mu m^2$. The bold line shows the mean order parameter value over $50$ realizations, and the shaded area shows one standard deviation from the mean.}
    \label{LargeArea}
\end{figure*}

In the derivation of our formula, we have approximated the spiking pattern as a Poisson distribution due to the irregularity of spike times in neural networks. One might expect that if the area of the neurons grows very large so that the magnitude of fluctuations is smaller and the dynamics of the stochastic model align very closely with the deterministic behavior, then the synchronization would not reach the steady state and would instead retain a synchronization value close to $1$ (i.e. completely synchronized). It might be tempting to think that one could ignore the fluctuations due to channel noise when the number of channels is extraordinarily large. Surprisingly, the answer to this is no. To examine this, we have considered an area of $300 \mu m^2$ (equivalent to $18,000$ sodium channels and $5,400$ potassium channels) whose results are shown in Fig. \ref{LargeArea}.

Despite the similarity of the the stochastic and deterministic dynamics (Fig.~\ref{LargeArea}), there are slight mismatches in spike timing due to the stochasticity of the Fox and Lu model. These slight mismatches accumulate over a lengthy period of time, and hence the order parameter decreases over time, albeit slowly. Even with the magnitude of the fluctuations from the channel noise being very small so that the spike timing is reasonably predictable, the degree of synchronization still decreases to the steady state value predicted by Eq.~(\ref{GS}). Our simulations and results suggest that channel noise should be accounted for in the modeling of real neuron synchronization regardless of the magnitude of fluctuations.

\subsection{Expected Time to Reach Steady State}
\label{ETTRSS}
\begin{figure*}
    \centering
    \includegraphics[width=0.75\textwidth]{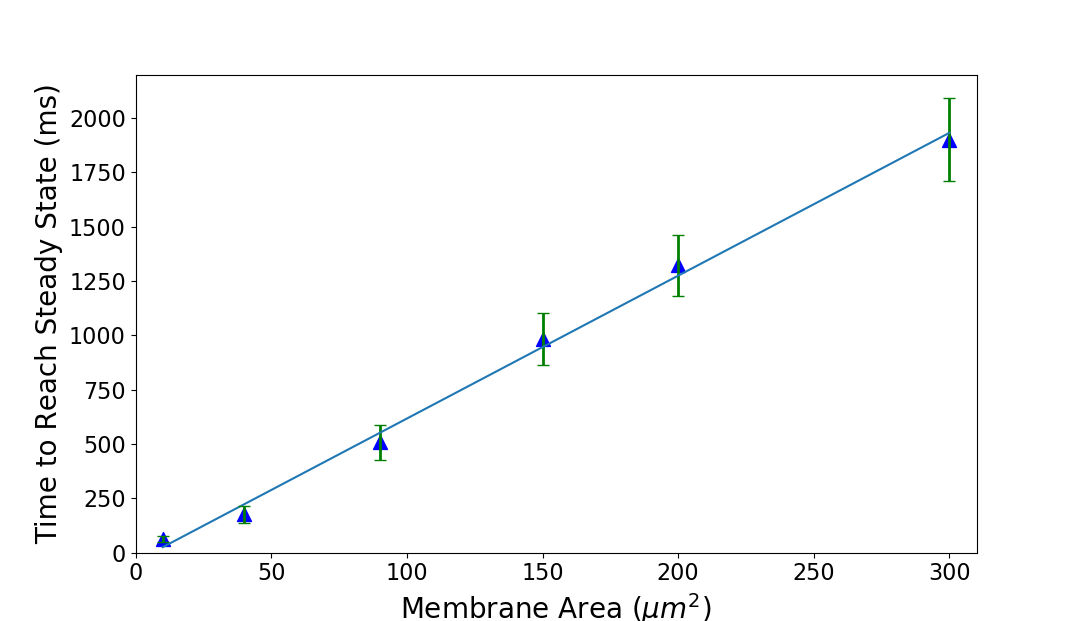}
    \caption{Expected time (triangles) for the mean order parameter to reach the steady state synchronization level. To calculate the expected time, we used the three neuron network in Fig. \ref{netw} with each neuron modeled by the Fox and Lu equations in Eqn. \ref{SHH}, injected each neuron with $10.0 \mu A/cm^2$, and measured the time for the mean order parameter over $50$ simulations to come within $0.02$ of the expected steady state based on Eqn. \ref{GS}. Vertical bars show $1$ standard deviation from the mean.}
    \label{ttss}
\end{figure*}

As we have shown, increasing membrane area does not change the expected steady state synchronization value, but instead increases the time it takes to reach that value. Here, we would like to quantify this observation. In our simulations, once the synchronization level gets near the expected value we calculated in Eq.~\ref{GS}, it hovers around that value. We consider the time to reach the steady state when our order parameter is within $0.02$ of the value calculated by Eqn. \ref{GS}. The results are shown in Fig. \ref{ttss}. Here we have considered the case of repetitive firing where we inject each neuron with a current of $10.0 \mu A/cm^2$. As suggested by Fig. \ref{ttss}, there is a linear relationship between the membrane area and the expected time to synchronization. This also means that in the limit of infinitely many channels where the dynamics become identical to the Hodgkin-Huxley model, the time to reach the steady state synchronization is ``infinite" in the sense that it never happens, which is in agreement with the order parameter of the Hodgkin-Huxley model never reaching a steady state value less than $1$. However, as long as the number of channels is finite, Fig. \ref{ttss} shows that the order parameter will eventually reach the expected steady state synchronization.

\section{Conclusion}
\label{conc}
In this paper, we have sought to examine the effects of channel noise on neural network synchronization. Because real neurons have finitely many channels, the stochastic opening and closing of these channels leads to fluctuations in the membrane voltage that are not accounted for in the deterministic Hodgkin-Huxley model. In order to account for these fluctuations, we used the Fox and Lu system size expansion model because (a) it is a highly accurate approximation to the gold standard (but computationally expensive) Markov Chain model, and (b) it is a far more computationally efficient model than the Markov Chain model~\cite{fox1994emergent,goldwyn2011stochastic}. We first looked at the relationship between firing frequency and input current in the presence of different magnitudes of channel noise. These simulations showed that channel noise had a larger effect on the firing rate in the absence of input current, but the effect was weakened as input current increased. We then looked at numerical simulations to qualitatively describe the effect of channel noise on neural network synchronization. We observed in Fig. \ref{DA} that (a) the neural networks hovered around a steady state synchronization level after some time, and (b) that increasing channel noise shortened the time it took to reach that synchronization level. In addition, we were able to derive a formula based on independent Poisson neurons to accurately estimate the long term expected steady state synchronization level. The key result is that even with a tiny amount of channel noise in coupled Hodgkin-Huxley equations, the steady state synchronization behaves identically to independent Poisson neurons. In addition, the derivation required $\lambda t$ to be large, where $\lambda$ is the firing rate. This requirement illustrates two features: that (a) as firing rate decreases (i.e. number of channels increases), the time to reach the steady state synchronization level increases, and (b) as firing rate increases, the steady state synchronization is reached faster. This observation was confirmed by the simulations shown in Fig. \ref{DA}. Our work suggests that despite the randomness within the model, channel noise causes neural networks to reach a steady state level of synchronization, and the steady state value only depends on the number of neurons in the network as suggested by Fig. \ref{SteadyState}.

We next considered two cases, one where the coupling constants were all different and one of them was much stronger than the others, and one where the membrane areas of each neuron were different but the coupling constants were the same. In both of these cases, we observed the same behavior of the synchronization level reaching the same value as predicted by independent Poisson neurons. We then considered the network in Fig. \ref{netw} where the membrane area was very large for each neuron, and each neuron was in the repetitive firing regime. The purpose of doing this was to observe the synchronization behavior when the stochastic dynamics are very close to the deterministic dynamics. Even when the dynamics are extremely similar, the small amount of channel noise causes a big change in the degree of synchronization. While three coupled Hodgkin-Huxley neurons (without noise) remain completely synchronized, the small amount of channel noise causes them to desynchronize as observed in Fig. \ref{LargeArea}. The reason for this is that the small variation in the timing of the spikes causes the phases of the neurons to gradually drift apart

Finally, we showed that there seems to be an approximately linear relationship between the membrane area and the expected time to reach the steady state synchronization level. This result shows that in the limit of infinitely many channels, the steady state synchronization we expect to reach from Eqn. \ref{GS} is never reached. This is in agreement with the simulation shown in Fig. \ref{LargeArea} where as the channel number grows to infinity, the stochastic dynamics converge to the deterministic dynamics, and the order parameter doesn't change for all time. However, for any finite number of channels, the simulation suggests that the expected steady state will be reached eventually. Ultimately, our paper strongly supports the notion that valuable insight can be gained by incorporating channel noise in the study of neural synchronization. In future work, we hope to compare our results obtained for the Fox and Lu model to other stochastic neuron models used to simulate channel noise.

\begin{acknowledgements}
We gratefully acknowledge support  by the U. S. Office of Naval Research (ONR) under Grant No. N00014-13-1-0205. We also wish to acknowledge Patrick Fitzsimmons for his help in deriving a formula for the steady state synchronization level.
\end{acknowledgements}

\section*{Appendix}
\label{AA}

The matrices used for numerical simulations and included in Eq.~(\ref{SHH}) are defined as:

\begin{widetext}
\begin{equation*}
A_K = 
\begin{bmatrix}
    -4\alpha_n & \beta_n & 0 & 0 & 0 \\
    4\alpha_n & -3\alpha_n - \beta_n & 2\beta_n & 0 & 0\\
   0 & 3\alpha_n & -2\alpha_n-2\beta_n & 3\beta_n & 0\\
     0 & 0 & 2\alpha_n &  -\alpha_n-3\beta_n & 4\beta_n\\
     0 & 0 & 0 & \alpha_n & -4\beta_n
\end{bmatrix}
\end{equation*}

\begin{equation*}
\resizebox{1.0\hsize}{!}{
$A_{Na}=
\begin{bmatrix}
    -3\alpha_m - \alpha_h & \beta_m & 0 & 0 & \beta_h & 0 & 0 & 0\\
    3\alpha_m & -2\alpha_m-\beta_m-\alpha_h & 2\beta_m & 0 & 0 & \beta_h & 0 & 0\\
    0 & 2\alpha_m & -\alpha_m-2\beta_m-\alpha_h & 3\beta_m & 0 & 0 & \beta_h & 0\\
    0 & 0 & \alpha_m & -3\beta_m-\alpha_h & 0 & 0 & 0 & \beta_h\\
    \alpha_h & 0 & 0 & 0 & -3\alpha_m-\beta_h & \beta_m & 0 & 0\\
    0 & \alpha_h & 0 & 0 & 3\alpha_m & -2\alpha_m-\beta_m-\beta_h & 2\beta_m & 0\\
    0 & 0 & \alpha_h & 0 & 0 & 2\alpha_m & -\alpha_m -2\beta_m -\beta_h & 3\beta_m\\
    0 & 0 & 0 & \alpha_h & 0 & 0 & \alpha_m & -3\beta_m-\beta_h
\end{bmatrix}$}
\end{equation*}

$S_K$ and $S_{Na}$ are the square root matrices of the following diffusion matrices:
\begin{equation*}
\resizebox{1.0\hsize}{!}{
$D_K =
\begin{bmatrix}
    4\alpha_n x_0 +\beta_n x_1 & -4\alpha_n x_0 -\beta_n x_1 & 0 & 0 & 0 \\
    -4\alpha_n x_0 -\beta_n x_1 & 4\alpha_n x_0 + \left(3\alpha_n+\beta_n \right)x_1 + 2\beta_n x_2 & -2\beta_n x_2-3\alpha_nx_1 & 0 & 0\\
   0 & -2\beta_n x_2-3\alpha_n x_1 & 3\alpha_n x_1 + (2\alpha_n+2\beta_n)x_2 + 3\beta_nx_3 & -3\beta_nx_3-2\alpha_nx_2 & 0\\
     0 & 0 & -3\beta_nx_3-2\alpha_nx_2 & 2\alpha_nx_2+(\alpha_n+3\beta_n)x_3 + 4\beta_nx_4 & -4\beta_nx_4-\alpha_nx_3\\
     0 & 0 & 0 & -4\beta_nx_4-\alpha_nx_3 & \alpha_nx_3+4\beta_nx_4
\end{bmatrix}$}
\end{equation*}

\begin{equation*}
\resizebox{1.0\hsize}{!}{
$D_{Na}= 
\begin{bmatrix}
    d_1 & -3\alpha_my_{00}-\beta_my_{10} & 0 & 0 & -\alpha_hy_{00}-\beta_hy_{01} & 0 & 0 & 0\\
    -3\alpha_my_{00}-\beta_my_{10} & d_2 & -2\alpha_my_{10}-2\beta_my_{20} & 0 & 0 & -\alpha_hy_{10}-\beta_hy_{11} & 0 & 0\\
    0 & -2\alpha_my_{10}-2\beta_my_{20} & d_3 & -\alpha_my_{20}-3\beta_my_{30} & 0 & 0 & -\alpha_hy_{20}-\beta_hy_{21} & 0 \\
    0 & 0 &  -\alpha_my_{20}-3\beta_my_{30} & d_4 & 0 & 0 & 0 & -\alpha_hy_{30}-\beta_hy_{31}\\
    -\alpha_hy_{00}-\beta_hy_{01} & 0 & 0 & 0 & d_5 & -3\alpha_my_{01}-\beta_my_{11} & 0 & 0\\
    0 & -\alpha_hy_{10}-\beta_hy_{11} & 0 & 0 & -3\alpha_my_{01}-\beta_my_{11} & d_6 & -2\alpha_my_{11}-2\beta_my_{21} & 0\\
    0 & 0 & -\alpha_hy_{20}-\beta_hy_{21} & 0 & 0 & -2\alpha_my_{11}-2\beta_my_{21} & d_7 & -\alpha_my_{21}-3\beta_my_{31} \\
    0 & 0 & 0 & -\alpha_hy_{30}-\beta_hy_{31} & 0 & 0 & -\alpha_my_{21}-3\beta_my_{31} & d_8    
\end{bmatrix}$}
\end{equation*}

\noindent and with diagonal entries:

\begin{flalign*}
d_1 &= (3\alpha_m + \alpha_h)y_{00} + \beta_my_{10} + \beta_hy_{01}&&\\
d_2 &= (\beta_m+2\alpha_m)y_{10}+2\beta_my_{20}+3\alpha_my_{00}+\alpha_hy_{10}+\beta_hy_{11}&&\\
d_3 &= (2\beta_m+\alpha_m)y_{20}+3\beta_my_{30}+2\alpha_my_{10}+\alpha_hy_{20}+\beta_hy_{21}&&\\
d_4 &= 3\beta_my_{30}+\alpha_my_{20}+\alpha_hy_{30}+\beta_hy_{31}&&\\
d_5 &= 3\alpha_my_{01}+\beta_my_{y11}+\beta_hy_{01}+\alpha_hy_{00}&&\\
d_6 &= (\beta_m+2\alpha_m)y_{11}+2\beta_my_{21}+3\alpha_my_{01}+\beta_hy_{11}+\alpha_hy_{10}&& \\
d_7 &= (2\beta_m+\alpha_m)y_{21}+3\beta_my_{31}+2\alpha_my_{11}+\beta_hy_{21}+\alpha_hy_{20}&& \\
d_8 &= 3\beta_my_{31}+\alpha_my_{21}+\beta_hy_{31}+\alpha_hy_{30}&&
\end{flalign*}
\end{widetext}

\bibliographystyle{apsrev4-1}
\bibliography{bibfile}

\end{document}